\documentclass[pra,aps,superscriptaddress,twocolumn]{revtex4}
\usepackage{graphicx}
\usepackage{amsmath}

\begin{document}

\title{An Introduction to Quantum Teleportation}

\author{Adam Miranowicz}
 \affiliation{Institute of Physics, Adam Mickiewicz University, 61-614
Pozna\'n, Poland}

\affiliation{CREST Research Team for Interacting Carrier
Electronics, School of Advanced Sciences,\\ Graduate University
for Advanced Studies (SOKENDAI), Hayama, Kanagawa 240-0193, Japan}

\author{Kiyoshi Tamaki}

\affiliation{CREST Research Team for Interacting Carrier
Electronics, School of Advanced Sciences,\\ Graduate University
for Advanced Studies (SOKENDAI), Hayama, Kanagawa 240-0193, Japan}

\date{\today; {\bf published} in {\em Math. Sciences  (Suri-Kagaku)}
{\bf 473}, 28--34, Nov. 2002} \pagestyle{plain}
\pagenumbering{arabic}

\begin{abstract}
Recent experiments confirm that quantum teleportation is possible
at least for states of photons and nuclear spins. The quantum
teleportation is not only a curious effect but a fundamental
protocol of quantum communication and quantum computing. The
principles of the quantum teleportation and the entanglement
swapping are explained, and physical realizations of teleportation
of optical and atomic states are discussed.
\end{abstract}

\maketitle \pagenumbering{arabic}

%%%%%%%%%%%%%%%%%%%%%%%%%%%%%%%%%%%%%%%%%%%%%%%%%%%%%%%%%%%%%%%%%%%%%%%%%%%%
\section{Introduction}

Teleportation is commonly understood as a fictional method for
disembodied transport: An object or person is disintegrated at one
place and it is perfectly reconstructed somewhere else. Thus,
teleportation can be compared to transmission of a
three-dimensional object using a kind of super fax machine which,
however, destroys the original object on scanning. In a sense, the
dokodemo-door (literally: door-to-anywhere) from the Japanese
Manga story ``Dora-e-mon'' can be considered as a teleporting
device (teleporter). In the movie ``Star Trek'', teleportation
serves as a standard transportation mean almost as common as
elevators. A Hollywood vision of possible dangers of non-perfect
teleportation has been created in the famous sci-fi movies ``The
Fly'' (1958,1986) and their sequels. Obviously, all these examples
sound completely unrealistic.

Until recently, physicists had ruled out teleportation because of
the implication of Heisenberg's uncertainty principle formulated
as the No Cloning Theorem, which prohibits making an exact copy of
an unknown quantum state. Yet, in 1993, an international team of
six scientists including Charles Bennett \cite{Ben93} demonstrated
that it is possible to transmit an unknown quantum state from one
place to another without propagation of the associated physical
object through the intervening space by way of a process called
the {\em quantum teleportation} (QT). The success of the first
experimental teleportation realized by Anton Zeilinger's group of
the University of Insbruck in 1997 \cite{Bou97} was the cover
story of many journals, including ``Scientific American''
\cite{Zei00}, and even newspapers all around the world. These
spectacular scientific achievements have caused much rumour.

Obviously, the underlying principles of the quantum teleportation
are fundamentally different from those of the dokodemo-door or the
``Star Trek'' teleporters of beaming people around. The phenomenon
that makes quantum teleportation possible is the {\em quantum
entanglement} also referred to as the Einstein-Podolsky-Rosen
(EPR) correlations \cite{Ein35}. Entanglement is a special
interrelationship between objects, in which measuring one object
instantly influences the other, even if the two are completely
isolated and separated from one another. For example, if two
photons come into contact with each other, they can become
entangled: The polarization of each photon is in a fuzzy,
undetermined state, yet the two photons have a precisely defined
interrelationship. If one photon is later measured by linear
polarizer to have, say, a vertical polarization, then the other
photon must collapse into the complementary state of horizontal
polarization. But if the entangled photon was measured by circular
polarizer, instead of the linear one, to have, say, a
right-circular polarization then the other photon had to collapse
into the complementary state of the left-circular polarization.
Thus, if one of the entangled photons is measured in any basis to
have a definite polarization, then the state of the other must be
exactly complementary to this polarization. It is so bizarre that
even Albert Einstein, who predicted this effect, considered it not
to be real and called it spooky \cite{Ein35}. An entangled state
of a system consisting of two subsystems cannot be described as a
product of the quantum states of these subsystems. In this sense,
the entangled system is considered inseparable and nonlocal.
Entanglement is usually manifested in systems consisting of a
small number of microscopic particles but, recently, it has also
been experimentally observed in macroscopic systems of $10^{12}$
atoms \cite{Jul01}, which sounds promising for further research in
teleportation of states of mesoscopic or even macroscopic objects.
Entanglement is one of the most profound features of quantum
mechanics having fundamental importance not only for quantum
teleportation but also for quantum computing and quantum
cryptography.

\section{Principles of quantum teleportation}

Here, we present the quantum teleportation protocol devised by
Bennett et al. \cite{Ben93} in 1993. This protocol is limited to
teleportation of states of a two-level quantum system, referred to
as the quantum bit or {\em qubit}. Nevertheless, generalization
for teleportation of states of multi-level or infinitely-level
systems is conceptually simple.

%%%%%%%%%%%%%%%%%%%%%%%%%%%%%%%%%%%%%%%%%%%%%%%%%%%%%%%%%%%%%%%%%%%%%%%%%%%%
%figure PRINCIPLE.
\begin{figure}[ht]
\vspace*{-5mm}\hspace*{-20mm}
\includegraphics[width=11cm]{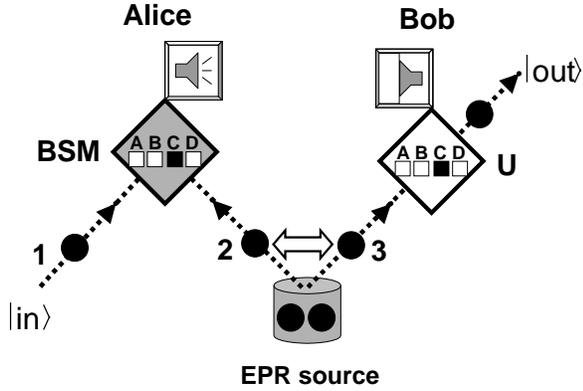} \vspace*{-20mm}%
\caption{Principles of quantum teleportation in the original
scheme of Bennett et al. \protect\cite{Ben93}:  Alice has a qubit
1 in an initial state $|in\rangle$, which she wants to teleport to
Bob. Besides Alice and Bob share an ancillary entangled (marked by
white arrow) pair of qubits (2 and 3) emitted by an
Einstein-Podolsky-Rosen (EPR) source. Alice performs a joint Bell
state measurement (BSM) on her qubit 1 and one of the ancillaries
(say qubit 2), projecting them onto one (say C as marked by black
box) of four orthogonal entangled states called the Bell states
(A, B, C, and D). Alice then sends to Bob the result of her
measurement by classical communication (symbolized here by
speakers). Dependent on this information, Bob does not change the
state of his qubit 3 (case A) or performs a unitary transformation
(U) on it (cases B-D) resulting in obtaining the output state
$|out\rangle$ of qubit 3 being exactly the same as the input state
$|in\rangle$ of the original qubit~1.}
\end{figure}

Qubit states to be teleported can be chosen as, for example,
the polarization states of single photon ($|\updownarrow\rangle$
and $|\leftrightarrow \rangle$, or $|+\rangle$ and $|-\rangle$),
the photon-number states of a cavity ($|0\rangle$ and
$|1\rangle$), the spin states of a spin-$\frac{1}{2}$ particle
(like electron) ($|\uparrow\rangle$ and $|\downarrow\rangle$),
ground and excited states of an atom or ion ($|g\rangle$ and
$|e\rangle$), or others. But it should be stressed that all these
realizations of qubits are mathematically equivalent. Thus, to
describe the principles of QT in this section, we use the standard
information notation of $|{0}\rangle$ and $|{1}\rangle$ for qubit
states.

The heart of teleportation are entangled qubits and the
measurement of their joint state performed in the basis of four
maximally entangled states, which are referred to as the {\em Bell
states} or the EPR states:
\begin{eqnarray}
|\Phi_A\rangle  &= \textstyle{\frac{1}{\sqrt{2}}}( |{0}\rangle|{1}
\rangle - |{1}\rangle|{0}\rangle)
\nonumber \\
|\Phi_B\rangle  &= \textstyle{\frac{1}{\sqrt{2}}}(|{0}\rangle|{1}
\rangle + |{1}\rangle|{0}\rangle)
\nonumber\\
|\Phi_C\rangle  &= \textstyle{\frac{1}{\sqrt{2}}}(|{0}\rangle|{0}
\rangle - |{1}\rangle|{1}\rangle)
\nonumber \\
|\Phi_D\rangle  &= \textstyle{\frac{1}{\sqrt{2}}}(|{0}\rangle|{0}
\rangle + |{1}\rangle|{1}\rangle) \label{Bell}
\end{eqnarray}
Quantum teleportation is a scheme by which the state of a qubit
can be transmitted from one place to another by classical
communication, provided that the sender (say Alice) and the
receiver (say Bob) have previously shared halves of two-qubit
entangled state. In detail, this is how it works: Assume that
Alice has a qubit 1 in the state $|in\rangle_1= a|{0}\rangle_1+
b|{1} \rangle_1$ with unknown amplitudes $a$ and $b$ normalized to
unity, $|a|^2+|b|^2=1$. In addition, Alice has qubit 2 initially
entangled to Bob's qubit 3 being in one of the Bell states, e.g.,
$|\Phi_A\rangle_{23}$, where subscripts 2 and 3 refer to relevant
qubits. The goal of teleportation is to transmit the state of
Alice's qubit 1 to Bob's qubit 3. The total initial state, given
by $|in\rangle_1 |\Phi_A \rangle_{23}$, can be rewritten in the
Bell basis as
\begin{eqnarray}
 &\textstyle{\frac{1}{2}}
 [|\Phi_A\rangle_{12} (a|{0}\rangle_3+b|{1}\rangle_3)+
 |\Phi_B\rangle_{12} (a|{0}\rangle_3-b|{1}\rangle_3)
\nonumber \\
 &- |\Phi_C\rangle_{12} (a|{1}\rangle_3+b|{0}\rangle_3)+
 |\Phi_D\rangle_{12}
 (a|{1}\rangle_3-b|{0}\rangle_3)]
\end{eqnarray}
simply by regrouping terms and omitting unimportant global phase
factor ($e^{i\pi}=-1$). Alice measures the joint state of qubits 1
and 2 in the Bell basis obtaining one of four possible results
\{A,B,C,D\}$\equiv$\{00,01,10,11\} (in classical bit notation).
The box BSM in figure 1 represents this Bell state measurement. No
matter what the two-qubit input state is, Alice's measurement
gives a uniformly distributed random two-bit classical result.
However, this measurement clarifies the difference between Alice's
initial qubit state and Bob's qubit: (A) If the measured qubits 1
and 2 are found to be in state $|\Phi_A\rangle_{12}$, then Alice's
classical output is $00$ and Bob's state is $|out\rangle_3= a|{0}
\rangle_3+b| {1} \rangle_3$, which is exactly the initial Alice's
state $|in\rangle_1$ without need to apply any additional
transformation ($U_A=I$); (B) If qubits 1 and 2 are found to be in
state $|\Phi_B\rangle_{12}$, then Alice's output is $01$ and Bob's
state is $|out_B\rangle_3= a|{0} \rangle_3-b|{1} \rangle_3$, which
differs from $|in\rangle_1$. However, a little thought shows that
by applying a simple phase flip $|x\rangle\mapsto(-1)^x |x\rangle$
($x=0,1$), which is realizable by the Pauli operator
$U_B=\sigma_z$, Bob gets state $|out\rangle_3= U_B|out_B\rangle_3$
being the exact copy of Alice's state $|in\rangle_1$. Similarly,
in cases C and D, Bob applies the proper unitary transformations
$U_C=-\sigma_x$ (a bit flip $|x\rangle\mapsto -|x\oplus 1\rangle$)
and $U_D=-i\sigma_y$ (a bit flip + phase flip $|x\rangle\mapsto
(-1)^{x+1} |x\oplus 1\rangle$) to obtain exactly the same state as
Alice's state $|in\rangle_1$. Thus, we conclude that to make a
successful teleportation, Alice has to inform Bob which of the
four states she measured, i.e., she must send him two bits of
classical information. Only then Bob can perform the correction
procedure by applying the appropriate unitary transformation $U$
($I$, $\sigma_z$, $-\sigma_x$ or $-i\sigma_y$) to recover the
initial Alice state. It is worth stressing that the transmission
of qubit states cannot be accomplished faster than light because
Bob must wait for Alice's measurement result to arrive before he
can recover the quantum state. Another important point is that
Alice by performing the Bell state measurement destroys the
initial state $|in\rangle_1$ of her photon. This loss of Alice's
state is the reason that QT does not violate the no-cloning
principle.

%%%%%%%%%%%%%%%%%%%%%%%%%%%%%%%%%%%%%%%%%%%%%%%%%%%%%%%%%%%%%%%%%%%%%%%%%%%%
%figure QSD
\begin{figure}[ht]
\vspace*{0mm}\hspace*{-15mm}
\includegraphics[width=10cm]{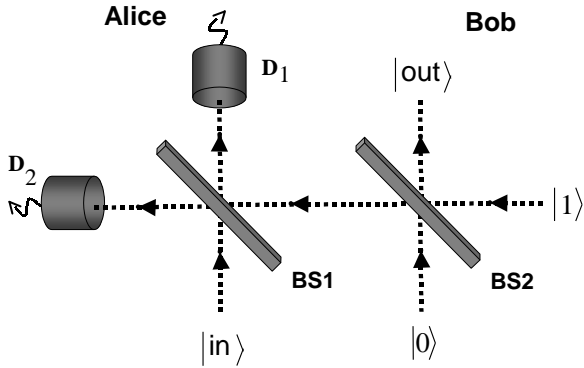} \vspace*{-22mm}%
\caption{Quantum teleportation and state truncation using quantum
scissors \protect\cite{Peg98}. If the inputs to the beam splitter
BS2 are single-photon state $|1\rangle$ and vacuum $|0\rangle$,
and Alice records one count at detector $D_1$ and none at $D_2$,
then the input state $|in\rangle=c_0|0\rangle +c_1|1\rangle
+c_2|2\rangle +\cdots$ is teleported and truncated to $|out\rangle
= c_0|0\rangle +c_1|1\rangle$ (with proper normalization). Thus,
if the input is the qubit state $|in\rangle=
c_0|0\rangle+c_1|1\rangle$ then it is just teleported (without
truncation). Bob, if informed by Alice about her measurement
result, knows that $|out\rangle=|in\rangle$. But it is equally
likely that Alice records one count at $D_2$ and none at $D_1$. In
this case, $|out\rangle$ would be phase flipped in comparison to
$|in\rangle$. Thus Bob, after receiving the information from Alice
about her measurement output, must perform on his state
$|out\rangle$ a simple unitary transformation of
$|x\rangle\mapsto(-1)^x |x\rangle$ to get the exact of copy of
Alice's state. If Alice records zero or two counts, then the
teleportation protocol fails. It follows that the probability of
successful teleportation is one in two.}
\end{figure}

So far, we have explained the teleportation for qubit states only.
As an example, we will discuss another protocol, which can be used
for teleportation of qubit states but also for state truncation.
In figure 2, we present a device proposed by David Pegg of
Griffith University, and Lee Phillips and Stephen M. Barnett of
the University of Strathclyde \cite{Peg98}, which is referred to
as the {\em quantum scissors}. In fact, this process can be
considered as the quantum teleportation since it is based on the
same principles as the original Bennett's scheme: (i)
entanglement, and so nonlocality, and (ii) the Bell-state
measurement (the projection postulate). The entangled state is
generated by Bob's beam splitter BS2 from the input states
$|0\rangle$ and $|1\rangle$. Please note that no light from input
qubit $|in\rangle$ reaches output qubit $|out\rangle$ so, indeed,
the process is a nonlocal phenomenon relying on quantum
entanglement. Alice implements her Bell-state measurement by beam
splitter BS1 and photon counters $D_1$ and $D_2$. And, as required
for teleportation, the original Alice's state $|in\rangle$ is
destroyed by her measurement. The first experiment with the
quantum scissors has been done by Alex Lvovsky's group of the
University of Konstanz \cite{Bab02}.

Bennett's protocol is deterministic (unconditional), which means
that every qubit state entering the setup can be teleported for
any output of Alice's measurement. The experimental protocols of
Akira Furusawa et al. carried out at the California Institute of
Technology \cite{Fur98} and Michael Nielsen et al. of the Los
Alamos National Laboratory \cite{Nie98} are also deterministic.
However, the experimental protocols of Zeilinger's group
\cite{Bou97} and Francesco De Martini's group of La Sapienza
University in Rome \cite{Bos98}, or also the quantum scissors are
only probabilistic (conditional): The teleportation is successful
conditional on appropriate results of Alice's measurement.

%%%%%%%%%%%%%%%%%%%%%%%%%%%%%%%%%%%%%%%%%%%%%%%%%%%%%%%%%%%%%%%%%%%%%%%%%%%%
\section{Atomic-state teleportation}

In the former section, we have described schemes for quantum
teleportation of optical qubit states since the first proposal
\cite{Ben93} and the majority of experimental implementations of
teleportation to date \cite{Bou97,Fur98,Bos98} were performed in
optical domain. However, from the practical point of view,
photonic qubits are not ideal for the long-term storage of quantum
information since they are very difficult to keep in certain place
as, e.g., light trapped in a cavity eventually leaks out. Thus, in
the quest for practical applications of quantum computers, the
teleportation of states of nuclear or atomic qubits attracts an
increasing interest to mention the spectacular experiment of
Nielsen et al. \cite{Nie98} of complete quantum teleportation of
states of nuclear qubits, where quantum state of carbon nucleus
was teleported to a hydrogen nucleus over interatomic distances
using nuclear magnetic resonance. Atoms (their electrons or
nuclei) are ideal for the long-term storage of quantum
information. Unfortunately, atoms move slowly and also interact
strongly with their environment and therefore they are not ideal
for the quantum information transfer at long distances. By
contrast, photonic states are ideal for the information
transmission but not the information storage. Here, we will
describe another interesting proposal of Peter Knight's group of
Imperial College \cite{Bos99} for teleportation of atomic states
over macroscopic distances. This scheme takes advantages both of
photons for transfer and atoms for storage of quantum information.

The crucial role in this teleportation protocol plays cavity
spontaneous photon leakage. It is a well accepted fact that
spontaneous decay of excited quantum systems is a mechanism of
their coherence loss (referred to as the decoherence) and
therefore usually plays a destructive role in quantum information
processing. However, Knight et al. have shown how detection of
decay can be used constructively not only for establishment of
entanglement but also for the complete quantum information
processing such as teleportation. This surprising result can be
understood by recalling the fact that a detected decay is a
measurement on the state of the system from which the decay
ensues.

%%%%%%%%%%%%%%%%%%%%%%%%%%%%%%%%%%%%%%%%%%%%%%%%%%%%%%%%%%%%%%%%%%%%%%%%%%%%
%figure DECAY
\begin{figure}[ht]
\vspace*{0mm}\hspace*{-10mm}
\includegraphics[width=9cm]{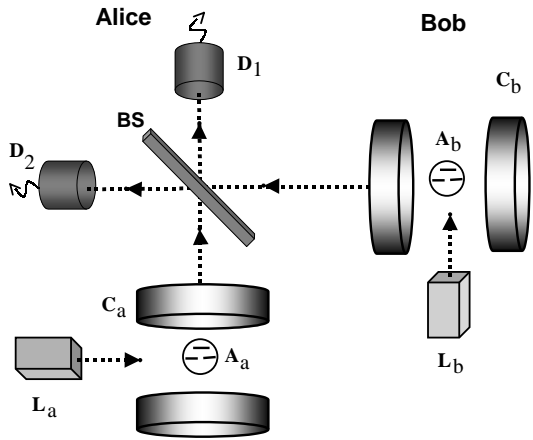} \vspace*{2mm}%
\caption{Atomic-state teleportation via decay
\protect\cite{Bos99}. The unknown state of Alice's atom $A_a$
trapped in her cavity $C_a$ can be teleported to Bob's atom $A_b$
trapped in a distant cavity $C_b$ by the joint detection by $D_1$
and $D_2$ of photons leaking from the cavities. At the preparation
stage, the atoms should be illuminated by lasers $L_a$ and $L_b$.
The state to be teleported is the internal state of an atom being
ideal for storing quantum information, while quantum information
is physically transferred from Alice to Bob via photonic states
being the excellent long-distance carriers of quantum information.
}
\end{figure}
%%%%%%%%%%%%%%%%%%%%%%%%%%%%%%%%%%%%%%%%%%%%%%%%%%%%%%%%%%%%%%%%%%%%%%%%%%%%
%figure LEVELS
\begin{figure}[ht]
\vspace*{-5mm}\hspace*{-5mm}
\includegraphics[width=9cm]{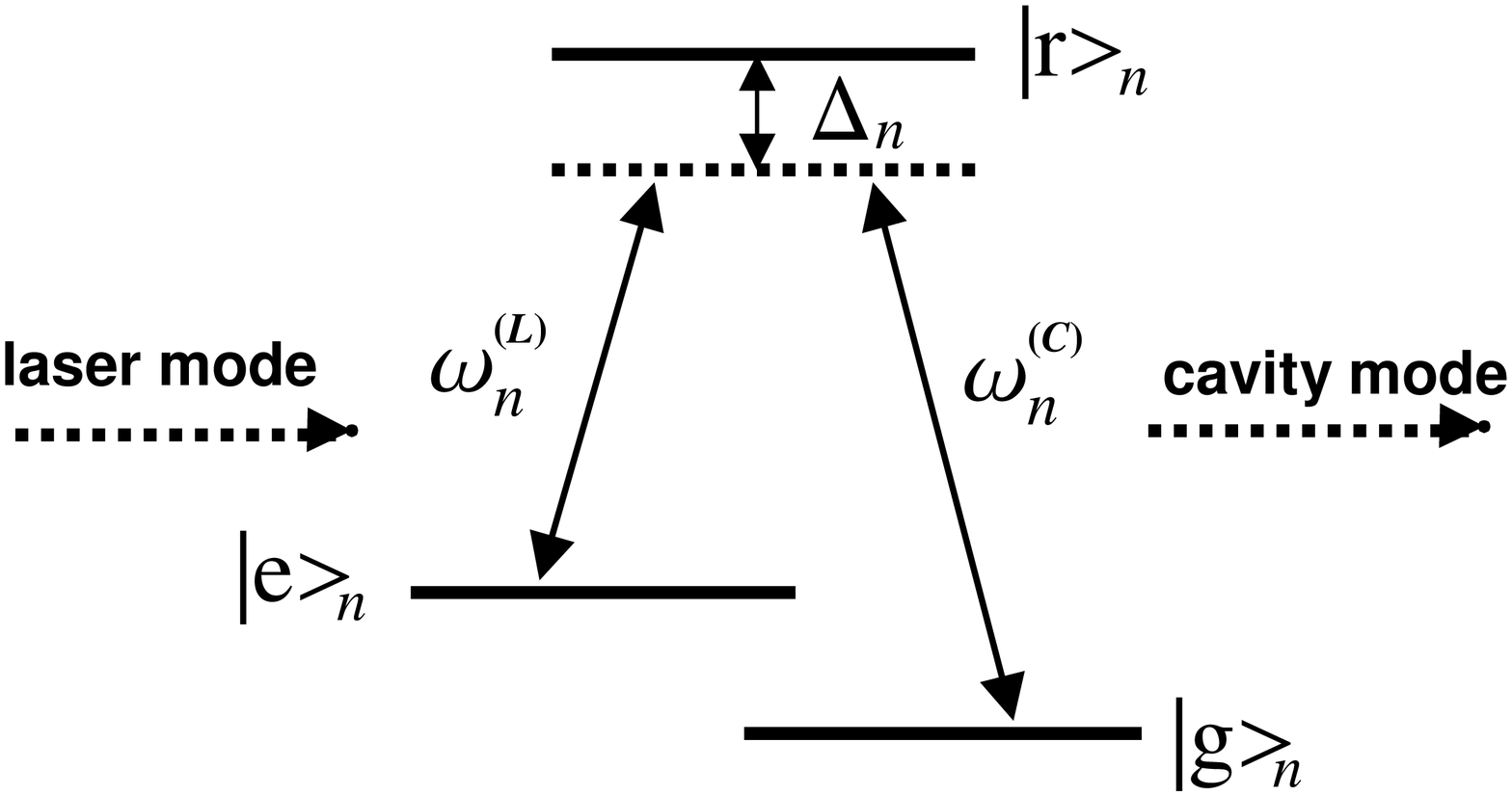} \vspace*{-15mm}%
\caption{Energy-level configuration of the atom trapped in Alice's
cavity ($n=a$) or Bob's cavity ($n=b$). Key: $|g\rangle_n$,
$|e\rangle_n$ -- atomic levels for the information storage;
$|r\rangle_n$ -- atomic excited level; $\omega^{(C)}_n$ --
frequency of the cavity field; $\omega^{(L)}_n$ -- frequency of
the classical laser field; $\Delta_n$ -- detuning.}
\end{figure}

Outline of the scheme is shown in figure 3. The setup consists of
two optical cavities: Alice's $C_a$ and Bob's $C_b$ tuned to the
same frequency $\omega^{(C)}_a=\omega^{(C)}_b$. Each cavity
contains a single trapped three-level atom ($A_a$ or $A_b$), which
is illuminated within a proper period of time by classical laser
field ($L_a$ or $L_b$). The atomic energy levels are depicted in
figure 4. By illuminating the atoms with the classical laser field
of frequency $\omega_n^{(L)}$, Alice (designated by subscript
$n=a$) and Bob ($n=b$) can drive the transition
$|e\rangle_n\leftrightarrow |r\rangle_n$. The other transition of
$|g\rangle_n\leftrightarrow |r\rangle_n$ is driven by the
quantized cavity field of frequency $\omega_n^{(C)}$. It is
important to assume that detunings $\Delta_n$ are large enough
such the upper levels $|r\rangle_n$ can effectively be decoupled
(so neglected) from the evolution of the lower levels. Thus, we
can assume that the quantum information is stored only in two
levels $|g\rangle_n$ and $|e\rangle_n$. Both Alice's and Bob's
cavities initially have no photons being described by vacuum state
$|0\rangle_n$, and Bob's atom is initially in state $|e\rangle$.
Alice does not know her atomic state, which is of the form
$|\psi\rangle_a=c |g\rangle_a + c' |e\rangle_a$ (with the unknown
coefficients $c$ and $c'$ such that $|c|^2+|c'|^2=1$). The main
task is to teleport the state $|\psi\rangle_a$ to Bob. First, as a
preparation of the state, Alice maps the atomic state
$|\psi\rangle_a$ on her cavity mode by illuminating the atom $A_a$
with the laser $L_a$ for a proper period of time. In the meantime,
Bob illuminates his atom $A_b$ with the laser $L_b$ for another
appropriate time period to generate an atom--cavity-field
entangled state $|\Psi\rangle_b=2^{-1/2} (|e\rangle_b|0\rangle_b
+i|g\rangle_b|1\rangle_b)$, where $|1\rangle_b$ and $|0\rangle_b$
stand for the cavity mode state with one or no photons,
respectively. Alice and Bob should synchronize their actions to
finish simultaneously the preparations of their states since
photons are leaking out from both the cavities. Those photons are
mixed on the 50–-50 beam splitter BS. Cavities are assumed to be
single-sided so that the only leakage of photons occur through the
sides of the cavities facing BS. The next step is the detection of
the photons, when Alice just waits for a finite time period for
click of the photon counter either $D_1$ or $D_2$. This joint
detection of photons leaking from distinct cavities $C_a$ or $C_b$
constitutes a measurement that enables a disembodied transfer of
quantum information from Alice's atom $A_a$ to Bob's atom $A_b$.
The cases, when Alice registers no clicks or two clicks, are
rejected as the failure of the teleportation. At the post
detection stage, Bob applies to the transferred state a proper
phase shift depending on whether detector $D_1$ or $D_2$ clicked.
This step corresponds to the unitary transformation $U$ described
in figure 1 and completes the teleportation protocol.

It is worth noting that the presented scheme, like the quantum
scissors, is probabilistic in the sense that the original state is
destroyed even if the teleportation fails, which is the case when
photon counters do not register one photon. However, the scheme
can be modified to a teleportation protocol with insurance by
entangling the initial Alice's atom $A_a$ with a reserve atom
$A_r$ also trapped in her cavity $C_a$ \cite{Bos99}.

%%%%%%%%%%%%%%%%%%%%%%%%%%%%%%%%%%%%%%%%%%%%%%%%%%%%%%%%%%%%%%%%%%%%%%%%%%%%
\section{Entanglement swapping}

Quantum teleportation of qubit states is one of the most
fundamental protocols of quantum communication. The generalized
version of the standard QT protocol is the {\em entanglement
swapping} \cite{Ben93,Zuk93}, where an entangled state of qubit is
teleported, as explained in figure 5. The QT and entanglement
swapping are essential parts of any quantum communication toolbox.
To show the similarities between the protocols we have used the
same symbols in figures 1 and 5. There are two equivalent ways to
interpret teleportation in the scheme: The state of qubit 1 is
teleported to qubit 3 or that of qubit 2 is teleported to qubit 4
after additional unitary transformation of the output states of
qubits 3 and 4, respectively. The first experiment demonstrating
the entanglement swapping was performed also by Zeilinger's group
\cite{Pan98}. They succeeded to entangle freely propagating
particles that never physically interacted with one another or
which have never been dynamically coupled by any other means. This
was probably the first direct experimental demonstration that
quantum entanglement requires the entangled particles neither to
come from a common source nor to have interacted in the past.

%%%%%%%%%%%%%%%%%%%%%%%%%%%%%%%%%%%%%%%%%%%%%%%%%%%%%%%%%%%%%%%%%%%%%%%%%%%%
%figure SWAPPING
\begin{figure}[ht]
\vspace*{0mm}\hspace*{-5mm}
\includegraphics[width=9cm]{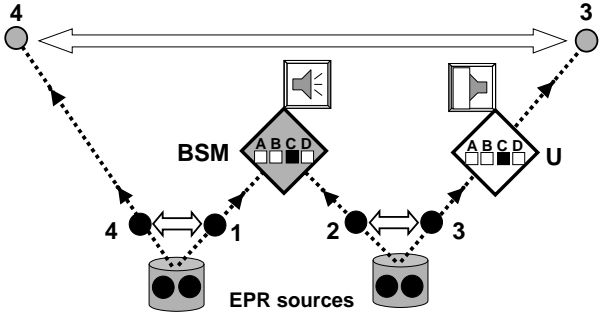} \vspace*{-15mm}%
\caption{Principles of the entanglement swapping: Application of
teleportation to entangle qubits that never interacted. Two EPR
sources produce two pairs of entangled qubits: 1--4 and 2--3
(marked by short arrows). Bell-state measurement (BSM), as
explained in figure 1, is performed on two qubits 1 and 2, one
from each pair. This measurement entangles outgoing qubits 3 and 4
(marked by long arrow).}
\end{figure}

\section{Entanglement Distillation}

As emphasized in all our discussions, EPR pair is an essential
resource for teleportation. So far, we have assumed that Alice and
Bob share the EPR states that can be accurately represented by for
instance $|\Phi_A\rangle$ in Eq.~(\ref{Bell}). Since Alice and Bob
cannot create the EPR pairs by classical communication, in order
to share the EPR pairs, distribution processes are required.
Suppose that Alice creates the EPR pairs and sends half of the
pairs to Bob. In realistic situations, due to the noise in
transmission channel or the imperfections of Alice's and Bob's
devises, the shared pairs are not the same as the ones represented
by $|\Phi_A\rangle$. Thus, for the QT protocol or other
applications, conversions from these imperfect pairs (less
entangled pairs) to the EPR pairs are needed. This process is
called the {\em entanglement distillation} and so far many
theoretical proposals on physical realization of the process have
been suggested \cite{distillation}. These proposals allow Alice
and Bob to distill the EPR pairs from less entangled pairs by
means of the operations performed separately by Alice and Bob, and
the classical communication. In 2001, the joint groups of Paul
Kwiat of Los Alamos National Laboratory and of Nicolas Gisin of
the University of Geneva \cite{Kwi01} have first demonstrated
experimental distillation of the EPR pairs by local filtering,
which is an operation individually applied to each distributed
pair. Recently, Nobuyuki Imoto's group at SOKENDAI \cite{Yam03}
successfully extracted an entangled photon pair from two
identically less entangled pairs by collective operations, which
are the operations applied to two distributed pairs. This is the
first experiment that involves the collective operations. These
experiments are a step towards experimental realizations of more
complicated applications in quantum information theory.

\vspace*{0.5cm}
\section{Perspectives and conclusions}

In Zeilinger's opinion, quantum teleportation between atoms
separated at macroscopic distances can experimentally be realized
within a few years and between molecules within a decade or so.
Nielsen's et al. \cite{Nie98} experimental teleportation of
nuclear-spin states although over microscopic distances is a good
prognostic. But, probably, the most spectacular and promising is
the experiment of Eugene Polzik's group of the University of
Aarhus \cite{Jul01} realizing the quantum entanglement between
macroscopic objects, i.e., a pair of caesium gas clouds containing
$10^{12}$ atoms each. Even though, the two samples were just
millimeters apart, they could in principle be entangled at much
longer distances. Entanglement of such large objects enables
`bulk' properties, like collective spin, to be teleported from one
gas cloud to another. Thus, the Polzik experiment, possibly, opens
the way for quantum teleportation between macroscopic atomic
objects.

The natural question arises what is the greatest difficulty in
teleporting people or other macroscopic objects. According to
quantum teleportation protocol, information from every tiny
particle in a person should be extracted, transferred to particles
elsewhere and assembled into an exact replica of the person. And
the problem is that human body is composed of about $10^{27}$ of
atoms and sending information about each individual atom state
(not only about collective properties as was done in Polzik's
experiment) would require, by applying the up-to-date technology,
time longer than that of the Universe. Thus, as emphasized by
scientists, anything even approximating quantum teleportation of
complex living beings, even bacteria or virus, is completely
beyond our technological capabilities.

Despite this unrealistic dream of new means of transportation,
quantum teleportation is not only a  trick but plays one of the
key roles in quantum information research of the last decade.
Without any doubt, it has already become an essential tool in
quantum computers \cite{Got99}, which in turn enable, if
constructed, super-fast calculations for simulation of the
Universe, weather forecasts or artificial intelligence. But
physicists must humbly admit that they only begin to understand
why, in fact, the quantum teleportation is possible in our quantum
world.

\vspace*{0mm} \noindent {\bf Acknowledgments.} It is our pleasure
to thank S.M. Barnett, G. Chimczak, A. Grudka, Y. Hirayama, N.
Imoto, M. Koashi, Yu-xi Liu, \c{S}. K. \"Ozdemir, R. Tana\'s, A.
W\'ojcik, and T. Yamamoto for stimulating discussions and
collaboration. We also thank Prof. Izumi Tsutsui for his kind
invitation to write this article.

%\vspace*{-2mm}
%%%%%%%%%%%%%%%%%%%%%%%%%%%%%%%%%%%%%%%%%%%%%%%%%%%%%%%%%%%%%%%%%%%%%%%%%%%%

\widetext
%%%%%%%%%%%%%%%%%%%%%%%%%%%%%%%%%%%%%%%%%%%%%%%%%%
\vspace{1mm} {\setlength{\fboxsep}{1pt}}
\begin{center}
\framebox{\parbox{0.75\columnwidth}{ \begin{center} published in
{\em Math. Sciences  (Suri-Kagaku)}  {\bf 473} (November 2002)
28--34
\end{center}}}
\end{center}

\end{document}